\documentstyle[12pt]{article} 
 
\begin{document} 
 \addtolength{\topmargin}{-1cm}
 \title{\bf Structural and Electronic Properties of Small Neutral 
          (MgO)$_n$ Clusters
          } 
 \author{ E. de la Puente$^1$, A. Aguado$^1$, A. Ayuela$^2$ and J.M. L\' opez$^1$ \\ \normalsize
 1. Departamento de F\'\i sica Te\'orica, Facultad de Ciencias,
  \\ \normalsize Universidad
  de Valladolid. 47011 Valladolid, Spain. \\ 
 \normalsize 2. Institut f\"ur Theoretische Physik, Technische Universit\"at Dresden, \\ \normalsize 01062 Dresden, Germany.} 

 \maketitle
 \begin{abstract}
 
{\em Ab initio} Perturbed Ion (PI)
calculations are reported for neutral stoichiometric $(MgO)_n$
($n \le 13$) clusters.
An extensive number of isomer structures was identified and studied.
For the isomers of $(MgO)_n$ ($n \le 7$) clusters, a full geometrical 
relaxation was considered.
Correlation corrections were included for all cluster sizes using 
the Coulomb-Hartree-Fock (CHF)
model proposed by Clementi. The results obtained
compare favourably to the experimental data and other previous theoretical
studies.  
Inclusion
of correlation is crucial in order to achieve a good description of these 
systems. 
We find an important number of new isomers which 
allows us to interpret the experimental magic numbers without  
the assumption of structures based on $(MgO)_3$ subunits.
Finally, as an electronic property, the variations in the
cluster ionization potential with the cluster size were 
studied and related to the structural isomer properties.

\end {abstract}

\bigskip

 

\newpage

\baselineskip 20pt
\section{Introduction}

Clusters provide a new state of aggregation from which the condensed matter
properties eventually emerge, being thus of formidable interest
in order to investigate the transition from
the molecular world to the solid state physics. Small clusters often present
substantial deviations in their physical and chemical properties when compared
both to the molecule and the bulk phase. Rationalizing the evolution of those
properties with the cluster size
is a major challenge to the nowadays science. 
As an example, for an
intermediate size cluster, the number of different isomers which may coexist in
a small energy range may be quite large. Though it is a very difficult task, 
a full characterization of those isomer configurations for each cluster
size would be highly desirable for a good understanding of cluster growing
processes.

In this paper we focus our interests in small stoichiometric $(MgO)_n$
clusters. These have received special attention in the last few years both from
experimentalists and theoreticians. Saunders\cite{Sau88,Sau89} published
mass-spectra and collision-induced-fragmentation (CIF) data for sputtered
$MgO$ cluster ions, and found enhanced stabilities for $(MgO)_n^+$ clusters with
$n=6$, $9$, $12$ and $15$. The results were interpreted there
in terms of $(MgO)_3$
subunits from which the clusters built up.
Ziemann and
Castleman\cite{Zie91a,Zie91b,Zie91c} performed experimental measurements by
using laser-ionization time-of-flight mass spectrometry\cite{Con88,Twu90}. They
found ``magic clusters'' at $n=2$, $4$, $6$, $9$, $12$ and $15$.
In order to explain the features observed in their mass spectra,
they performed theoretical calculations by using the rigid ion and the
polarizable ion shell models\cite{Mar83,Die84,Phi91}. Their main conclusion
was that the clusters form compact cubic structures similar to pieces of the
$MgO$ crystal lattice. Self consistent calculations on these clusters were
performed by Moukouri and Noguera\cite{Mou92,Mou93} with the use of a 
semiempirical tight binding approach. 
The necessity of going
beyond a fully ionic model like that of pair potential interactions was there
remarked. 
Recio {\em et al}\cite{Rec93a,Rec93b} reported {\em ab initio} molecular 
orbital calculations on neutral clusters of $MgO$ containing up to 26 atoms.
These calculations included correlation effects in the clusters with $n=$
1-6 molecules.
Their results were in agreement with
Saunders' hypothesis, in the sense that structures based on the $(MgO)_3$
subunit were preferred over the cubic like ones.
Very recently, Malliavin {\em et al}\cite{Mal97} have performed calculations
on $(MgO)_n$ ($n \le 6$) clusters using the DMOL method. The geometries
obtained in that work are in total agreement with those presented by Recio
{\em et al}.

Each particular work from the above mentioned ones represented, in principle,
an 
improvement over the preceding investigations. 
Pair potential models provided
the first theoretical attempt to identify and classify different isomers and
calculate 
the energy differences between them. Nevertheless, some selected set of
empirical parameters for $Mg^{q+}$ and $O^{q-}$ ($q=$ 1,2) ions
has to be used as input, and that
set is the same for ions which are in nonequivalent cluster sites.
The studies by Moukouri and Noguera added self consistency to the calculations,
but they were not free of empirical parameters. Finally, the first
{\em ab initio} study on these systems was carried out by
Recio {\em et al}. Several approximations had to be done in their calculations,
however, in order to maintain the computational time at reasonable values:
(a) Large
geometrical distortions in their considered isomers were not allowed, as the
optimization process was carried out just by varying the nearest neighbour
distance $R$ (for ringlike structures, the stacking distance between rings was
also independently optimized) and, 
for some structures, even this value was kept fixed;
(b)
Several other isomers, which may be important in a more complete 
characterization
of the different isomer structures 
adopted by small $(MgO)_n$ clusters, were not considered;
(c) Correlation corrections could be included
(at the MP2 level) only in clusters containing up to 12 atoms.

The present work undertakes a new
extensive and systematic {\em ab initio} study of $(MgO)_n$ clusters with
$n$ up to 13. We have used the
{\em ab initio} Perturbed Ion (aiPI) model\cite{Lua90a}, which is based on the
Theory of Electronic Separability (TES)\cite{Huz71,Lua87} and the 
{\em ab initio} Model Potential (aiMP) approach of Huzinaga {\em et al}\cite{Huz87}, within the Restricted Hartree-Fock (RHF) approximation. The model has 
been used by our group in several studies on $(NaCl)_n$\cite{Ayuel}
and
$(NaI)_n$\cite{Agu96a} clusters, and also in a global study of alkali halide
clusters\cite{Agu96b}.
Although we do not
expect to obtain accurate results for the only partially ionic
$MgO$ molecule, our
main concern is by far not the molecule, but intermediate size clusters where a
full range of different isomer structures may be studied. The solid $MgO$ has
been excellently described by the aiPI model\cite{Lua90b}, and precise results
are achieved for $(MgO)_n$ clusters from $n=3$ on. 

Our calculations represent a
major advance with respect to pair potential or semiempirical methods. The
ion-cluster consistency achieved in the calculations results in a different
{\em ab initio} description of each nonequivalent ion in the cluster, without
requiring the use of any empirical parameter\cite{Ayuel,Agu96a,Agu96b}. 
On the other hand, they 
represent an alternative description to the molecular orbital
models, and complete the results obtained by Recio {\em et al} in the
following
aspects:
(a)
For $(MgO)_n$ ($n \le 7$)
clusters, a full geometrical relaxation of the different isomers 
has been considered;
(b) We have studied a more complete set of isomers;
(c) Correlation corrections 
have been included for all cluster sizes (up to 26 atoms).
Finally, 
some electronic properties as ionization potentials have been also considered.

Results presented here are also aimed to assist in the
interpretation of possible future experimental investigations on these clusters.
Renewed interest in isomer geometries has recently emerged because of the
drift tube experimental studies, which, by measuring the mobility of cluster
ions through an inert buffer gas under the influence of a weak electric field,
provide valuable information about the cluster geometries\cite{Hel91,Jar95,Mai96}. However, we have no knowledge on experimental studies of this kind on
$(MgO)_n$ clusters. 

The rest of the paper is structured in several sections. In the next one, we
give a brief r\'esum\'e of the aiPI model for the study of clusters
(the interested reader can obtain more details in 
our previous works\cite{Ayuel,Agu96a,Agu96b}),
and some computational details. Section III deals with the principal structural
and electronic results of the present study. Conclusions are given in
the last section.

\section{Computational Method}

The {\em ab initio} perturbed ion model\cite{Lua90a} was originally designed
for the description of ionic solids\cite{Lua92}, and subsequently adapted to
the study of clusters in our group\cite{Ayuel,Agu96a,Agu96b}. Its theoretical
foundation lies in the theory of electronic separability\cite{McW94} for
weakly overlapping groups\cite{Fra92}, and its practical implementation in
the Hartree-Fock (HF) version of the TES\cite{Huz71,Lua87}.
The HF equations of the
cluster are solved in localized Fock spaces, by breaking the cluster wave
function into local nearly orthogonal group functions (ionic in nature in our
case). When the self-consistent process (see details below) finishes,
the outputs are the total
cluster energy $E_{clus}$ and a set of localized wave functions for
each geometrically nonequivalent ion in the cluster. The cluster energy can be
written as a sum of ionic additive energies\cite{Ayuel,Agu96a,Agu96b}:
\begin{equation}
 E_{clus}=\sum_{R=1}^N E_{add}^R,
\end{equation}
where the sum runs over all ions in the cluster, and the contribution of each
particular ion to the total cluster energy ($E_{add}^R$) can be expressed in
turn as a
sum of intraionic (net) and interionic contributions:
\begin{equation}
 E_{add}^R=E_{net}^R+\frac{1}{2}\sum_{S(\not= R)}E_{int}^{RS}=E_{net}^R+
\frac{1}{2}E_{int}^R.
\end{equation}
Once $E_{clus}$ has been obtained, the binding energy per molecule of the
$(MgO)_n$ cluster with respect to the dissociation process $(MgO)_n \rightarrow
nMg^{2+} + nO^{-} + ne^- + nE_{bind}$, is given by:
\begin{equation}
E_{bind}=-\frac{1}{n}[E_{clus} - nE_0(Mg^{2+}) - nE_0(O^{-})],
\end{equation}
where $E_0(Mg^{2+})$ and $E_0(O^{-})$ are the energies of the $Mg^{2+}$ and
$O^{-}$ free ions, respectively.

The localized nature of the aiPI procedure has some advantages over the usual
molecular orbital models. As in weakly overlapping systems the correlation 
energy correction is almost intraionic in nature (being therefore a sum of
contributions from each ion), the localized cluster-consistent ionic wave
functions may be used to attain good estimations of this correction. In this 
paper, the correlation energy correction is obtained through Clementi's
Coulomb-Hartree-Fock method\cite{Cle65,Cha89}.
Besides, it also allows the development of computationally efficient codes\cite{Lua93} which make use of the large multi-zeta basis sets of Clementi and
Roetti\cite{Cle74} for the description of the ions. At this respect, our
optimizations have been performed using basis sets (5s4p) for $Mg^{2+}$
and (6s4p) for $O^{2-}$, respectively. Inclusion of diffuse basis functions
has been checked and shown unnecessary. A general discussion on the 
election of the most appropriate
basis set within the aiPI model has been given 
elsewhere\cite{Agu96b}.

A few comments regarding the weak-overlap assumption and the description of the
$O^{2-}$ anions are worth of mention. $O^{2-}$ is an unstable anion in free
space, so a basic problem in an ionic description of $MgO$ clusters is the
correct description of the $O^{2-}$ wave function. Lua\~na {\em et al} have
shown how the $O^{2-}$ anion is stabilized by the action of the crystal
environment in solid $MgO$\cite{Lua90b}. Specifically, the 2p oxygen orbital
experiences a large contraction in the lattice, which makes it stable as an
embedded anion. The resulting $O^{2-}$ wave function leads to a diamagnetic
susceptibility and an electron kinetic energy increase in quantitative agreement
with the experiment. Thus, the quality of that wave function is asserted.
An analysis of the PI results reveals that the orbital contraction is mainly due
to the action of the projection operators. The same conclusion has been achieved
for the $MgO$ clusters studied here. The projection operator supplies two major
effects\cite{Fra92}: On one hand, it tries to maintain the strong-orthogonality
hypothesis, being therefore responsible for the orbital contraction. On the 
other hand, it has a well-defined physical meaning, namely, its expectation
value coincides exactly with the overlap energy if we assume that this overlap
is ``weak''. By ``weak'' it is meant that the products of two overlap integrals
$({\bf S}_{AB})_{ij}({\bf S}_{CD})_{km}$ with $A \neq B$ and $C \neq D$ are
negligible ($({\bf S}_{AB})_{ij} = < \phi_i^A \mid \phi_j^B >$, where $\phi_i^A$
is the {\em ith} orbital of the ion $A$).
This corresponds to the following
truncation of the L\"owdin expansion\cite{Low56} for
${\bf S}^{-1}$:
\begin{equation}
{\bf S}^{-1} \approx {\bf I} - {\bf T} \; \; \; \; \; \; \; \; \; \; \; \; \; \; \; \; \; \; ({\bf T} = {\bf S} - {\bf I}).
\end{equation} 
Within this context, the hypothesis of weakly overlapping ions does not imply
that the overlap contribution to the cluster energy is negligible. What is
neglected is those contributions to the overlap energy coming
from second and higher orders in the L\"owdin expansion. 
The overlap energies of the $O^{2-}$ anions 
in the 
$(MgO)_3$ clusters
are just $\sim 0.5 eV$ larger than the corresponding value in the solid
(12.84 eV for $O^{2-}:MgO$), and they are decreasing
functions of the cluster size. 
The assumption of weakly overlapping ions is good in the solid (the results are
exceptional), and we consider that such an small increase in the overlap
energies does not invalid that assumption for cluster studies.
Our basis set for oxygen anion is flexible
enough because by adding diffuse basis functions the cluster energy does not
lower anymore,
nor the
projection operator is more effective in projecting the frozen orbitals out of
the active space 
(we would like to remark that an extensive set of different basis
functions was considered in test calculations on $(MgO)_2$ and 
$(MgO)_6$ isomers. The ones
selected gave the lowest cluster energies and achieved saturation in the value
of the overlap energies).

We have
used the following self-consistent method:
for a given distribution of the ions
forming the cluster, we consider one of them as the active ion R (for
instance, a particular oxygen anion), and solve the Self-Consistent-Field 
equations
for anion R
in the field of the remaining ions, which are considered frozen at this
stage. The solution
obtained is transferred to all the anions equivalent 
to anion R, that is, to the anions
which have equivalent positions in the cluster. Next, we take a non
equivalent oxygen anion (anion S) as the active ion and repeat the same process.
Evidently, since anions S are not equivalent to anions R, the energy
eigenvalues and wave functions of electrons in anions S are 
different from those of anions R.
We continue this process 
in the same way until all the inequivalent anions have been
exhausted. 
The same procedure is then followed for the magnesium cations. The process
just described is a PI cycle. We iterate the PI cycles until
convergence in the total energy of the cluster is achieved.
 
As input geometries we considered cubic structures resembling pieces of
bulk $MgO$ crystal, ring structures (mainly hexagonal, but also some octagonal 
and decagonal), mixed structures made up usually of a regular isomer with
a molecule attached to it, and some more open structures like the truncated
octahedron for $n=12$ or the wurtzite piece for $n=7$. These trial geometries
have been either taken from the geometries considered in 
pair potential calculations on $MgO$ and alkali
halide clusters or intuitively guessed with the experience gathered in our
previous works on alkali halide clusters. The input geometries have been
fully relaxed (that is, the total cluster energy has been minimized with 
respect to variations in all the $(3N-6)$ independent variables, where $N$
is the number of ions) for $(MgO)_n$ ($n \le 7$) clusters, at the HF level.
A simplex downhill algorithm has been used in these calculations\cite{Nel65,Wil91}.
For clusters with $n=$8-13 this procedure became computationally quite 
expensive 
and we turned to geometrical optimizations (also at the HF level)
with respect to a limited number of relevant parameters.
Cuboid structures have been relaxed with respect to a single parameter,
the first-neighbour cation-anion distance, whereas in the ring structures
the cation and anion
distances to the center of the ring, and the distance between rings have been
independently varied.
For the mixed structures, the
part of the regular structure which is nearest (first and second
neighbours) to the attached molecule is relaxed together with the
molecule, and the rest of the isomer is kept fixed at the geometrical
configuration with the minimum energy
found for that isomer without the molecule attached. The CHF
correlation correction is then introduced for all the studied isomers at the
equilibrium geometries found at the HF level. These geometries are further
scaled (without shape modifications) until the minimum CHF energy is
achieved. We do not expect, from the results obtained by 
Recio {\em et al}\cite{Rec93a},
that inclusion of correlation substantially modifies the isomers shape. 
We
have carried out these scaling calculations mainly 
because they can affect differently to ringlike and cubiclike structures
(and therefore to the energetic isomer ordering) and also
in order to assess
its influence on the averaged cluster bond length, and its evolution
towards the bulk value.

\section {Results.}

\subsection {Lowest energy structures and isomers.} 

Results concerning the structural properties of small neutral $(MgO)_n$
clusters, obtained by following the method explained in the previous section,
are shown in figure 1. Small and large spheres are used to represent $Mg^{}$
and $O^{}$ ions, respectively. The most stable CHF structure (first isomer)
is shown on the left side for each cluster size. The rest of the structures
are the low-lying isomers obtained in the calculation. For clusters with 3, 4
and 5 molecules we show two different views of some isomers. The numbers given
below each isomer are the total cluster energy differences with respect to the
ground state (denoted GS in the figure), at the CHF (first row) and HF
(second row) levels.

For small clusters sizes ($n=$ 2-6) only two isomers have been considered. The 
ground state geometries in the HF calculations are mainly rings: hexagonal
for $n=$ 3, decagonal for $n=$ 5 and a stacking of two hexagonal 
rings for $n=$ 6.
However for $n=$ 4 the GS is a cube. The inclusion of correlation does not
change the GS structures except for $n=$ 5, where a distorted cube with a $MgO$ 
molecule attached to it becomes energetically more stable. These results
are in total agreement with the results of Recio {\em et al}\cite{Rec93b} 
and Malliavin {\em et al}\cite{Mal97}.
However, the isomers obtained in our calculations  have large geometrical
distortions compared to those obtained by Recio {\em et al}, 
mainly those which have
non-ringlike configurations: the first excited isomer for $n=$ 3 was obtained
after a full relaxation of a planar piece of the $MgO$ bulk lattice; 
this planar 
piece, with no distortions, is also the first excited isomer in Recio {\em et
el} 
results, but we can see in fig. 1 that the final geometry is greatly distorted
with respect to the perfect bulk structure. This kind of deformations has been
also obtained in lithium halide [22] and sodium iodide [23] clusters.
The above results show that the PI calculations agree with those 
of other more ``popular'' {\em ab-initio} calculations: 
GAUSSIAN\cite{Rec93a,Rec93b} and DMOL\cite{Mal97}. 
That means that the PI method is a good 
candidate for obtaining the ground state and other geometrical and electronic
properties of clusters with ionic bonding like the $(MgO)_n$ clusters 
considered in this paper.

For the clusters with $n=$ 7-13 we have considered a larger number of initial
geometries in our calculations. Overall, the ground state geometries obtained
in the HF calculations are the same that those 
obtained by Recio {\em et al}. We appreciate differences only for 
those cluster sizes in which the ground state structures that we
obtain were not considered by Recio {\em et al} in their calculations
($n=$ 11,12,13).

The inclusion of correlation in the calculations changes the GS
geometries for $n=$ 7 and 8. These results can not be compared with
the calculations performed by Recio {\em et al} because they did not include
the correlation contribution in this range of sizes; neither with the
calculations of Malliavin {\em et al} because they did not perform any
calculations in this size range.
For $n=$ 7 the HF ground state isomer is the
wurtzite piece whereas the CHF ground state is the bulk-like $(MgO)_6$
piece with a $MgO$ molecule attached to it. For $n=$ 8 the HFGS
is a stacking of two octagonal rings whereas the CHFGS is a
mixed structure formed by ``joining'' the most stable $(MgO)_6$ and
$(MgO)_4$ isomers. The last structure is also the second isomer 
in the HF results. For $n=$ 9 the ground state isomer is a stacking of 
three hexagonal rings, the same as in the HF calculations; the 
other isomers considered are a
piece of $MgO$ bulk and the $(MgO)_8$ HFGS with a molecule attached
to it. For $n=$ 10 the GS geometry is a piece of the $MgO$ lattice. 
The other isomers considered in our calculations are, in order of
increasing energy: a stacking of two decagonal rings, a piece 
formed by ``joining'' two stackings of two hexagonal rings by 
a face, and a piece of bulk $MgO$ lattice with a molecule attached
to it (attaching a molecule to the $(MgO)_9$ GS leads to an isomer which lies
higher in energy). 
For $n=$ 11 the GS isomer is a mixed 
structure composed by a stacking of two hexagonal rings plus a bulk-like 
part, and the
other isomers considered are either from mixed type or bulk-like. 
A truncated octahedron is obtained as the GS for $n=$
12. This isomer has been also 
found to be specially
stable in $(NaI)_{12}$ clusters\cite{Agu96a}.
It was not considered in ref.\cite{Rec93b}.
Finally, a mixed structure composed by a piece of the $MgO$ lattice and an
octagon is found as the $(MgO)_{13}$ ground state. 
Our second isomer is the defect-cuboid structure, which was
the single isomer tried in ref.\cite{Rec93b}. 
A point 
which is worth of mention from these two final cluster sizes is that there is
the possibility to obtain different isomers with the same geometrical shape
but with cations and anions interchanged. The electrostatic interaction tends
to stabilize energetically anions\cite{Ayuel,Agu96a}, so those isomers with a 
central oxygen are more stable.

We compare now the preceding results to the results 
obtained with the PI model for 
alkali-halide $(AX)_n$ clusters\cite{Ayuel,Agu96a,Agu96b}.
In our previous works we found evidence of a clear-cut separation between the
structural properties of different alkali-halide clusters. $(KX)_n$ and
$(RbX)_n$ clusters showed almost from the very beginning a preference to
adopt minimum energy structures which are fragments of the bulk lattice.
However, $(LiX)_n$ clusters adopted preferentially more open structures.
$(NaX)_n$ showed an intermediate behaviour. 
Our structural results for $(MgO)_n$
clusters show, in outline, large similarities with those of $(LiX)_n$ or
$(NaI)_n$ clusters.
These are the least ionic compounds in the alkali-halide series. Rounded
geometries had been proposed by Twu {\em et al}\cite{Twu90} in those cases
showing a decreased ionic bonding character. 
Then, our minimum energy isomers
are consistent with a reduced ionic character in $(MgO)_n$ clusters. This
non-fully ionic character was also pointed out in the analysis of Recio
{\em et al}\cite{Rec93b}. The discussion there
was based on a comparison with the classical results of Ziemann 
{\em et al}\cite{Zie91a}, who performed pair potential
calculations by using two different (but both fully ionic) models, with
different nominal charges assigned to the ions. The calculations for
$(Mg^{+1}O^{-1})_n$ clusters revealed mainly cubic structures, whereas in
$(Mg^{+2}O^{-2})_n$ calculations a set of much more open structures was
obtained. As the $(Mg^{+1}O^{-1})_n$ calculations explained better the 
experimental mass spectra, Ziemann {\em et al} concluded that pieces of $MgO$
bulk lattice are obtained for small $(MgO)_n$ clusters. Both our calculations
and those of refs.\cite{Rec93b,Mal97} 
obtain results which are in between these two
limit cases, so nominal charges between 
$\vert1 \vert$ and $\vert2 \vert$ should be
ascribed to $Mg/O$ ions in $classical$ pair potential simulations. However,
our calculations show that a description in terms of weakly overlapping
$Mg^{2+}$ and $O^{2-}$ groups is completely appropriate provided that all the 
relevant quantum-mechanical interactions (overlap included)
are accounted for properly 
and a full ion-cluster consistency is achieved.
It is worth of stressing that such a description does not enter in conflict
with a possible assignment of fractional charges to each ion. A population
analysis is best carried out directly in terms of the total density, following
the Bader's scheme for deriving atomic properties from the topology of the
charge density\cite{Bad90}. That has been recently done in PI calculations on
solids\cite{Mar97}, and fractional ionic charges have been derived for a
variety of ionic crystals. 

From the
study of small $(MgO)_n$ ($n=$ 2-13) clusters we conclude that, in the earliest
stages of cluster growing,
the bulk lattice fragments are not the GS isomers
except for some particular values of $n$. For the rest of
cluster sizes other (more open) structures are preferred. 
Thus, the convergence to the bulk structure is not reached yet in this 
size range 
(the solid $MgO$ crystallizes in the f.c.c. structure\cite{Ash76}). 

We can appreciate in fig.1 for $n=$ 6, 9 and 12
(and also for other cluster sizes), that the inclusion of
correlation enhances the stability of
the hexagonal isomers over the cubic ones. Though in the size range covered
open structures are preferred over the cubic ones, eventually there will be a 
cluster size where the cubic structures become energetically more stable, and
this specific value of $n$ will depend on the inclusion or not of correlation
in the calculations. The above observation suggests that this
``critical'' $n$ will be larger upon inclusion of correlation, that is, 
pieces of 
the $MgO$ bulk lattice will appear as GS isomers for values of $n$ which are
larger in CHF calculations than in 
HF calculations.

The magic numbers
for small $(MgO)_n^+$ clusters emerging from mass spectra experiments have been
explained by Saunders\cite{Sau88} and by Recio {\em et al}\cite{Rec93b} as
resulting from specially enhanced structures based on the stacking of $(MgO)_3$
subunits. Our results show that, at least for neutral $(MgO)_n$
clusters (and also for ionized $(MgO)_n^+$ clusters if the vertical
approximation holds\cite{Rec93b}), this conclusion is not universal. 
The truncated octahedron
obtained as the ground state isomer 
for $n=$ 12 in our calculations indicates that this interpretation should be
taken with some care.

\subsection {Inter-ionic distances and binding energies.}

In figure 2 we present the average interionic CHF distance $d$ between
$Mg^{}$ and $O^{}$ 
nearest neighbours for the
ground state geometry 
as a function of the cluster size $n$.
$d$ presents an irregular behaviour
as a function of $n$, due to the fact that the ground state geometries
for the different values of $n$ do not correspond to the
same
type of structural family.
The points that correspond to the same structural family have been joined 
by the two dashed lines, and it can be
appreciated that the variation of $d$ is smoother
for each structural family. The value of $d$ increases with the 
number of ions in the cluster. 
The
cubic clusters have an interatomic distance larger than the hexagonal ones.
The isomers which resemble pieces of the
bulk $MgO$ lattice are
the important set in order to compare with the
bulk value.
In the $MgO$ crystal $d=$ 2.106 \AA\cite{Ash76}, and we
obtain a value of 
$d=$ 1.999 \AA $~$ for the GS isomer with $n=$ 10. 
The saturation value is not
achieved yet, but the trend is the correct one. The contraction of the 
distances induced by the inclusion of correlation is more important in
cubiclike pieces than in ringlike ones:
the average effect in cubic structures is a contraction of approximately
4 \%, while in hexagonal structures the contraction is only of some 2 \%.
The inclusion of correlation is necessary
in order to obtain
quantitative agreement with the experimental bulk value in the limit of $n$
large. 

In Figure 3, we represent $E_{bind}$
as a function of $n$, both for the HF and CHF calculations.
The general trend is an increase of the
binding energy with $n$ in both cases. 
Superposed to that general trend, 
especially stable clusters are predicted for certain values of
$n$ (``magic numbers''); 
these are identified in the figures as maxima or pronounced changes in
the slope.
We observe that both type of calculations do not predict the same set of
magic numbers. The HF curve is quite smooth for small values of $n$, and maxima
are observed only at $n=$ 9, 12 (pronounced slope changes are not observed).
The inclusion of correlation effects (CHF curve) is crucial to achieve agreement
with the experimental results\cite{Sau88,Zie91a}, which predict magic
clusters at $n=$ 2, 4, 6, 9, 12, 15... The maxima at $n=$ 6, 9, 12
are clear, and a pronounced slope change is appreciable at $n=$ 4. The abundance
maximum obtained experimentally at $n=$ 2 should be 
probably related to the
singly-ionized stoichiometric clusters and not to the neutrals\cite{Rec93b}.

\subsection {Cluster ionization potentials.}

In this subsection we analyse the variation of the cluster ionization potential
(IP) with the cluster size. The output of the PI calculation contains a
fully cluster-consistent wave function for each inequivalent ion in the cluster.
In our model,
the electron must be extracted from a specific localized anionic
orbital when the cluster is ionized. The vertical IP of the cluster is 
calculated as
the smallest binding energy of a 2p anionic
electron (Koopmans' approximation).
The results are given in figure 4, where we have plotted the eigenvalues of the
2p orbitals of $O^{2-}$ anions as a function of $n$ for
the most stable structure of each cluster size in the HF (fig.4a)
and in the CHF (fig.4b) calculations.
We appreciate a band of eigenvalues for each cluster size
because the anions can occupy inequivalent positions in the cluster.
These band gaps are indicative of the different reactivity of each nonequivalent
cluster site, they can
determine the
preferred adsorption sites, and could be helpful in photoionization spectroscopy
studies.

The dashed lines, which join the lowest 2p energy eigenvalues for each cluster
size, indicate the variation of the HF (4a) and CHF (4b) ionization potentials
with the cluster size.
The variation of the vertical cluster ionization potential with $n$ is related
to the relative stabilities of neutral and singly-ionized stoichiometric
clusters\cite{Rec93b}: When there is a maximum in the IP for a cluster size
$n$, the ionized structure is less stable than the neutral one
against loss of a $MgO$ monomer;
the opposite occurs at the minima in the IP curve. 
For the HF results, a set
of minima at $n=$ 4, 7, and 11 and maxima at $n=$ 3, 5, 8, and 10 is observed,
in agreement with the results of Recio {\em et al}. 
The difference is
that they obtain a maximum at $n=$ 12; but the results are not directly
comparable
for these cluster sizes ($n=$ 12, 13), because the isomers that we obtain as the
GS were not
considered in
their calculations.
This pattern changes upon inclusion of correlation
effects (see CHF curve):
minima are found at sizes $n=$ 5, 7, and 11, and maxima at $n=$
3, 6, and 10. These changes are directly related to the change in the GS isomer
for $n=$ 5.
We can give a clear geometrical
interpretation to the minima in the light of the localized picture provided
by the PI model. At the minima ($n=$ 5, 7, and 11), mixed structures were found
as the most stable isomers (fig.1). The least bound anionic p-electron in those
clusters corresponds always to that oxygen with the smallest number of
neighbouring ions.
That specific oxygen is less stabilized than the
other oxygens in the cluster because of the reduced electrostatic interaction 
for those cluster sites. An electron
will be removed from it more easily than for the rest of cluster sizes (for 
which the GS isomer is more ``compact'').
Such a clear geometrical interpretation of the maxima in the IP's is not 
apparent for the small size range covered in this work:
the opposite situation (maxima in the IP for those ground states with the
highest symmetries) is not always observed.
As the main contribution to the anionic binding energy is
the Madelung potential, and this is the largest for the bulklike 
structures,
we expect
maxima to appear precisely in compact cubic structures for larger cluster sizes.

In simple metallic clusters, the variations in the cluster ionization potential
are correlated with the magic numbers, indicating that the magic numbers
are electronic in nature\cite{Hee87}. 
There is not such a definite
connection in ionic clusters, and the variations observed in the cluster 
ionization potential are related mainly to structural features.  
From our discussion, the existence of such a connection
between electronic (IP's) and geometrical properties in $(MgO)_n$ clusters
becomes clear.
The question of the interconnection between electronic, energetic,
and geometrical cluster properties 
has been
more thoroughly addressed 
in ionic alkali-halide materials\cite{Agu96b}. 

\section {Conclusions}

{\em Ab initio} Perturbed Ion (PI) calculations have been carried out in order
to investigate the structural, energetic and some electronic properties of
small neutral stoichiometric $(MgO)_n$ clusters. The set of studied isomers 
has been more complete than in previous calculations, and some of the not
previously considered isomers have resulted to be either the ground state 
structure or a low-lying isomer for several cluster sizes. 
Specifically, we have shown that the interpretation of the magic numbers of
small $(MgO)_n$ clusters in terms of stackings of $(MgO)_3$ subunits is not
universal, because a truncated octahedron is more stable for $n=$ 12 than the
corresponding hexagonal isomer.
General structural
distortions have been considered for $(MgO)_n$ $(n \le 7)$ clusters, and these
deformations have been shown to be really important in those isomers with the
lowest symmetries. 
Correlation corrections have been included for all cluster
sizes with the Clementi's Coulomb-Hartree-Fock (CHF) method. Their inclusion
has been very important in the determination of some ground state isomers 
and of
the energy differences between isomers for each cluster size.
Specifically, at
$n=$ 5, 7, 8, and 10 different ground state structures are obtained after
accounting for correlation effects. Comparison with previous {\em ab initio}
calculations has been made and the agreement has been good whenever the 
comparisons have been possible.

The evolution of the average inter-ionic distance with the cluster size has 
been studied. Clusters from the same structural family show a smooth variation
of their inter-ionic distances with $n$. The correlation-induced distance
contractions are the largest for those isomers which are fragments of the $MgO$
bulk lattice. Energetically, however, the correlation correction stabilizes
more the hexagonal isomers, so the transition towards bulk structures will
take place at larger cluster sizes than HF calculations would predict. 
This result contrasts with the situation encountered in alkali halide clusters,
where the inclusion of correlation induces a larger energetical stabilization
of the cuboid isomers\cite{Agu96b,Och94}.
The
$(MgO)_n$ magic numbers have been identified as maxima or pronounced slope
changes in the binding energy {\em versus} 
cluster size curve. They are $n=$ 4, 6, 9,
and 12, in agreement with the experimental results $n=$ 2, 4, 6, 9, 12, 15, ...
($n=$ 2 has been considered to be due to a larger enhancement of the 
singly-ionized $(MgO)_2^+$ species). Inclusion of correlation has also 
shown to be
completely necessary in order to reproduce the correct (experimental)
magic numbers.

The interconnection between electronic and structural cluster
properties has been
studied by considering the variations in the 
cluster ionization potential (IP) with
the cluster size $n$. 
For the
minima in the IP, which occur for the lowest symmetry ground state isomers
at $n=$ 5, 7, and 11, 
a clear geometrical interpretation can be given: in those isomers, there is
always an oxygen anion with a number of neighbouring ions notably reduced, from
which is relatively easy (compared to the rest of cluster sizes) to remove an
electron and create a singly-ionized isomer.
Such a definite connection has not been found for the
maxima in the size range considered in this work.
The corresponding opposite situation of maxima in the IP emerging for all those
ground state isomers with the highest symmetries is not observed.
The increased classical electrostatic (Madelung) energy
of isomers resembling pieces of the $MgO$ crystal explains the 
maximum observed at $n=$ 10.
Then,
maxima for larger cluster 
sizes should be expected to appear whenever the ground state becomes a 
compact fragment of the $MgO$ bulk lattice.

$\;$

$\;$

$\;$

$\;$

$\;$

{\bf ACKNOWLEDGEMENTS}. This work has been supported by DGICYT (Grant
PB95-D720-C02-01). One of us (A. Aguado) acknowledges a predoctoral Grant from
Junta de Castilla y Le\'on.

\newpage

{\bf Captions of figures}

$\;$

$\;$

Figure 1. {Isomer geometries for $(MgO)_n$ clusters. The most stable
CHF structure (first isomer) is shown on the left side. Total energy differences (in eV)
with respect to the most stable structure are
given for each isomer (first row: CHF; second row: HF). $Mg^{2+}$, small spheres; $O^{2-}$, large spheres.
For $n=3-5$, two different views of some isomers are provided.

$\;$

$\;$

Figure 2. Averaged nearest neighbour Mg-O distance for the
CHF ground state structures of Figure 1.
The two lines join cubic and hexagonal clusters.

$\;$

$\;$

Figure 3. Binding energy per molecule as a function of the cluster size.

$\;$

$\;$

Figure 4. Orbital energies (with opposite sign) of the 2p levels of $O^{2-}$
anions
as a function of the cluster size for the ground state structures of $(MgO)_n$
clusters. The dashed line joins the vertical cluster ionization potentials.
(a) HF results; (b) CHF results.

\pagebreak

\end{document}